\documentclass[RNAAS]{aastex631}

\begin{document}

\title{Recent updates of gas-phase chemical reactions and molecular lines of SiS in CLOUDY}

\correspondingauthor{Gargi Shaw}
\email{gargishaw@gmail.com}

\author[0000-0003-4615-8009]{Gargi Shaw}
\affiliation{Department of Astronomy and Astrophysics\\ Tata Institute of Fundamental Research\\  
Mumbai 400005, India}

\author[0000-0003-4503-6333]{Gary Ferland}
\affiliation{Department of Physics and Astronomy\\ University of Kentucky\\
Lexington, KY 40506, USA}

\author[0000-0002-8823-0606]{M. Chatzikos}
\affiliation{Department of Physics and Astronomy\\ University of Kentucky\\
Lexington, KY 40506, USA}

\begin{abstract}
Here we present our current update of CLOUDY on gas-phase chemical reactions for the formation and destruction of 
the SiS molecule, its energy levels, and collisional rate coefficients with H$_2$, H, and He over a wide range of temperatures. As a result, henceforth the spectral synthesis code CLOUDY predicts SiS line intensities and column densities for various astrophysical environments. 

\end{abstract}



\section{Introduction} \label{sec:intro}
The spectroscopic simulation code CLOUDY simulates physical conditions in NLTE astrophysical
plasma over a wide range of physical parameters (temperature, density, radiation field etc.) using \textit{ab initio}  microphysics, and predicts the resulting column density and spectra
over the entire electromagnetic range. Details about CLOUDY can be found in \citet{{2013RMxAA..49..137F},{2017RMxAA..53..385F},{2012ApJ...746...78G},{2005ApJ...624..794S},{2017ApJ...843..149S},{2020RNAAS...4...78S},{2022ApJ...934...53S}}\footnote{\url{https://nublado.org}}.

Silicon sulfide (SiS) has been observed in massive star formation regions \citep{1988ApJ...324..544Z} and 
circumstellar envelopes around 
evolved stars \citep{1993AJ....105..576B}. 
Its rich rotational spectrum can be observed using ALMA. Motivated by this, we recently updated the 
spectral synthesis code CLOUDY to predict the column density and rotational line intensities of SiS in various 
astrophysical environments. This work is basically an extension of our previous work \citep{2022ApJ...934...53S} 
 which aimed to predict more molecular lines with better accuracy.

With our git repository management, 
GitLab hosting infrastructure and continuous autonomous
testing, we aim to move the Cloudy project to a more
rapid release schedule, something like the Ubuntu project. 
As part of this development cycle, we aim to announce
improved physics in various publications.
The advance in the chemistry described here gives
Cloudy greater power to interpret molecular
line observations.

\section{Calculations and Results}
All numerical calculations presented here are performed using the development version of the spectral simulation code CLOUDY, 
last described by \citet{{2017RMxAA..53..385F},{2022ApJ...934...53S}}.

The density and line intensity of SiS lines depend on the chemical reaction rates and the internal structure,
together with Einstein's radiation coefficients and collisional rate coefficients. We include energy levels, 
Einstein's radiation coefficients and collisional rate coefficients with H$_2$ for the SiS using LAMDA 
database \citep{2005A&A...432..369S}. It includes energy levels up to $\it{J}$=40. There are 40 radiative and 820 collisional transitions covering a temperature range of 10 to 2000 K. \citet{ANUSURI2019e01647} quantum dynamically calculated rotational excitations of SiS with H. It also includes 11 transitions over the temperature range of 5 to 300 K. However, CLOUDY uses collisional data in the LAMDA format, which uses deexcitation rate coefficients. Hence, we calculate the deexcitation rate coefficients from the excitation rate coefficients following the formula,
\begin{math} 
\Gamma_{ul}= \Gamma_{lu}\times(g_l/g_u)\times \exp(h\nu/kT)
\end{math}. For the given temperature range, H-collisional rate coefficients are larger than the 
H$_2$-collisional rate coefficients. However, collisions with H$_2$ are more important than 
the collisions with H because of its larger abundance where SiS is present.
Further, we include collisional rate coefficients with He following \citet{{2007A&A...472.1037V},{2008JPhB...41o5702T}} for the lowest four rotational levels for a temperature range of 40-1400 K. This range of temperature is set by the calculations of \citet{{2007A&A...472.1037V},{2008JPhB...41o5702T}}.

We include 32 formation and 10 destruction reactions for SiS from the UMIST Database for Astrochemistry, UDfA \citep[RATE12;][]{{1997A&AS..121..139M},{2013A&A...550A..36M}}. In addition, we include four more reactions, SiH + S $\rightarrow$ SiS + H, 
SiH + S$_2$ $\rightarrow$ SiS + SH, Si + SO$_2$ $\rightarrow$ SiO + SO, and Si + SO$_2$ $\rightarrow$ SiS + O$_2$ from \citet{2018ApJ...862...38Z}. All four of these reactions have a constant rate coefficient, 1$\times$10$^{-10}$ cm$^3$s$^{-1}$. Reactions, Si + HS$\rightarrow$ SiS + H, S + SiN $\rightarrow$ SiS + N, 
S + SiH$_2$ $\rightarrow$ SiS+ H + H, S + HCSi $\rightarrow$ SiS + CH, 
Si + S$_2$ $\rightarrow$ SiS + S are taken from \citet{1998A&A...330..676W}. Whereas, 
Si + H$_2$S $\rightarrow$ SiS + H$_2$ , H$_2$ + SiS$^+$ $\rightarrow$ HSiS$^+$ + H, 
and S+ SiH$_2^+$ $\rightarrow$ HSiS$^+$ + H are taken from \citet {2021SciA....7.7003D} and the Kinetic Database for Astrochemistry, 
respectively\footnote{\url{https://kida.astrochem-tools.org/}}.
We found that among these reactions, Si + HS $\rightarrow$ SiS + H plays a very significant role in the formation of SiS.

As a test, we first model a generic H~II region and a PDR calculation to study the abundances of SiS for similar  environments. Our model is described by  \citet{2022ApJ...934...53S}, and the input script of this model, ``h2$_{-}$orion$_{-}$hii$_{-}$pdr.in'', is publicly available with the \textsc{cloudy} download under the directory \texttt{tsuite}. In a nutshell, the H~II region starts at a distance of 10$^{17.45}$~cm away 
from the ionizing star and extends into the PDR and dense molecular region
up to $A_{\rm V}$ = 1000 mag. The hydrogen density at the ionized face is 10$^4$ cm$^{-3}$ with the gas-phase abundances similar to the average gas-phase abundances in the Orion Nebula. 
\begin{figure}[ht!]
\plotone{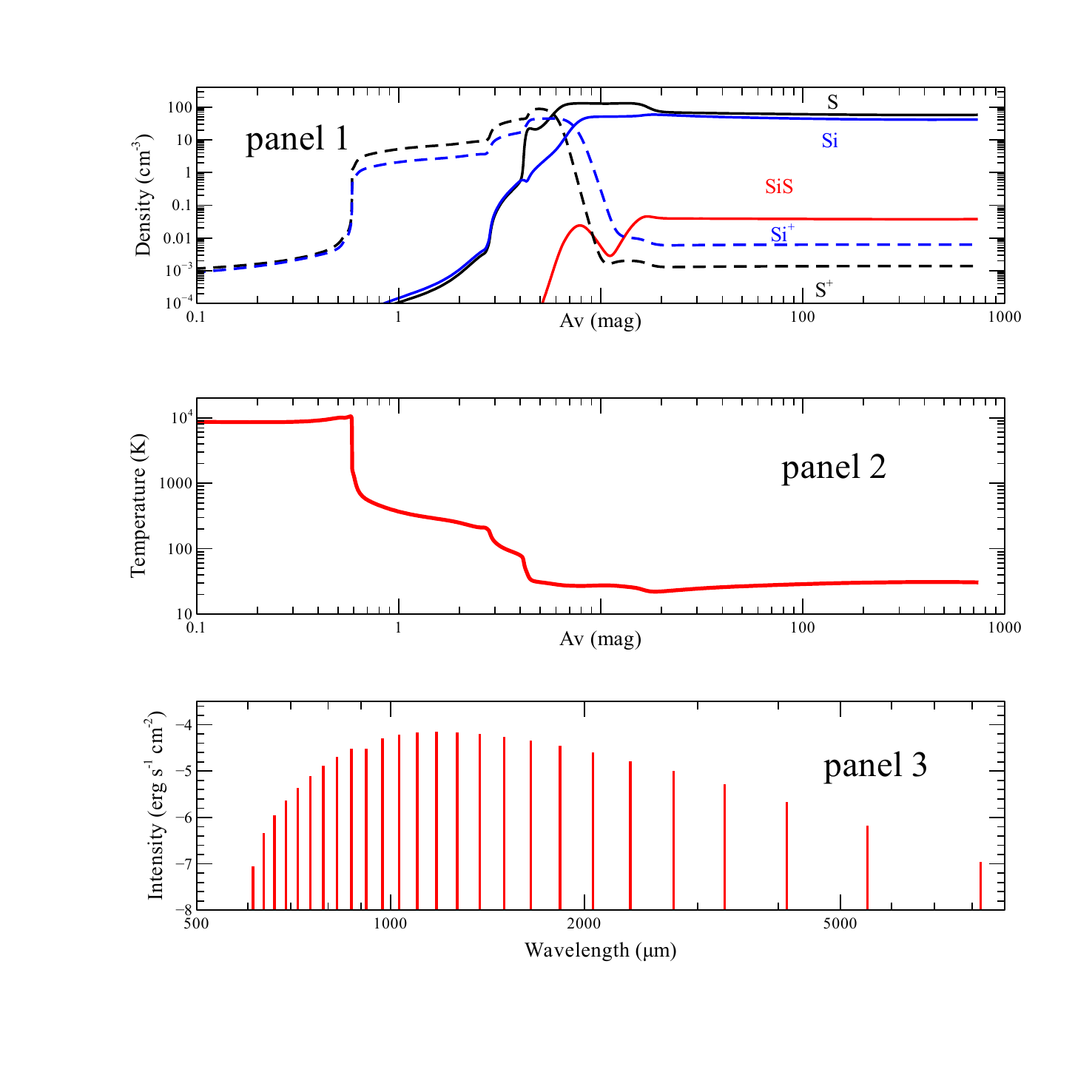}
\caption{Panel 1: Variation of SiS density as a function of $A_{\rm V}$ for an H~II and PDR model (``h2$_{-}$orion$_{-}$hii$_{-}$pdr.in'' from the \textsc{cloudy} download under the directory \texttt{tsuite}.) Panel 2: Variation of temperature across the cloud. Panel 3: Model predicted intensities of various SiS rotational lines. 
\label{fig:fig1}}
\end{figure}
Panel 1 of Figure \ref{fig:fig1} shows the variation of density of SiS, S, Si, S$^+$, and Si$^+$ across the cloud as a function of Av. 
The red, blue, and black solid lines represent the abundances of SiS, Si, and S. Similarly, blue and black dashed lines represent the abundances 
of Si$^+$ and S$^+$. Panel 2 of Figure \ref{fig:fig1} shows the temperature across the same cloud. Panel 3 shows the predicted intensities 
of various rotational lines of SiS that can be observed by ALMA. It is clear from the plot that SiS forms deep in the cloud (Av $>$ 10) where 
the temperature is lower (25 to 35 K). Our model predicted column density of SiS is 10$^{15.76}$ cm$^{-2}$. These advances will be part of 
the 2023 release of CLOUDY.

In future, we will include higher vibrational energy levels and rovibrational transitions of SiS to CLOUDY to make it applicable to JWST observations.

\begin{acknowledgments}
GS acknowledges WOS-A grant from the Department of Science and Technology (SR/WOS-A/PM-2/2021).
GJF acknowledges support by NSF (1816537, 1910687), NASA (ATP 17-ATP17-0141, 19-ATP19-0188), and STScI (HST-AR- 15018 and HST-GO-16196.003-A).
MC acknowledges support by NSF (1910687), and NASA (19-ATP19-0188, 22-ADAP22-0139).
\end{acknowledgments}

%





\bibliography{SiS_update}{}
\bibliographystyle{aasjournal}



\end{document}